# Charting the landscape of nefarious uses of generative artificial intelligence for online election interference

by Emilio Ferrara

**Abstract**
Generative artificial intelligence (GenAI) and large language models (LLMs) pose significant risks, particularly in the realm of online election interference. This paper explores the nefarious applications of GenAI, highlighting their potential to disrupt democratic processes through deepfakes, botnets, targeted misinformation campaigns, and synthetic identities. By examining recent case studies and public incidents, we illustrate how malicious actors exploit these technologies to try influencing voter behavior, spread disinformation, and undermine public trust in electoral systems. This paper also discusses the societal implications of these threats, emphasizing the urgent need for robust mitigation strategies and international cooperation to safeguard democratic integrity.

**Contents**

## Introduction

The advent of GenAI is revolutionizing fields from human-AI teaming to creative content generation, enabling new forms of human-computer interaction and augmenting human capabilities in unprecedented ways (Brown, *et al.*, 2020; OpenAI, *et al.*, 2023; Rahwan, *et al.*, 2019). However, these advancements come with significant risks: here we focus on the problem of interference with democracy (Bovet and Makse, 2019; Ozmen Garibay, *et al.*, 2023) within the broader context of AI ethics (European Commission. High-Level Expert Group on Artificial Intelligence, 2019; Mittelstadt, *et al.*, 2016). The primary objective of this study is to investigate the nefarious applications of GenAI in online election interference, exploring how these technologies can disrupt democratic processes. We aim to provide a comprehensive overview of the threats posed by GenAI, illustrating dual nature paradox that this technology poses, by means of a literature review and analysis to identify key trends and case studies (Barrett, *et al.*, 2023).

Online election interference refers to the manipulation of digital information and communication technologies to influence the outcome of elections (Kreps and Kriner, 2023; Tenove, *et al.*, 2018). This can include activities such as spreading misinformation, creating fake news, and orchestrating social media campaigns to sway public opinion. The introduction of GenAI into this arena has magnified these threats by enabling more sophisticated and scalable forms of interference (Goldstein, *et al.*, 2023; Mozes, *et al.*, 2023; Shoaib, *et al.*, 2023). Generative AI, with its ability to create realistic text, images, and videos, presents new opportunities for malicious actors to deceive and manipulate the public. For example, deepfake technology can generate realistic videos of political figures saying or doing things they never did, eroding public trust in authentic information sources; see Figure 1 (Barrett, *et al.*, 2023; Gabriel, *et al.*, 2024; Kenthapadi, *et al.*, 2023; Lorenz, *et al.*, 2023; Seymour, *et al.*, 2023). Moreover, the widespread use of AI-powered botnets can amplify misinformation and create the illusion of widespread support or opposition to certain political ideas, further distorting the democratic process (Coeckelbergh, 2023; Kreps and Kriner, 2023; Unver, 2018; Yang and Menczer, 2023).

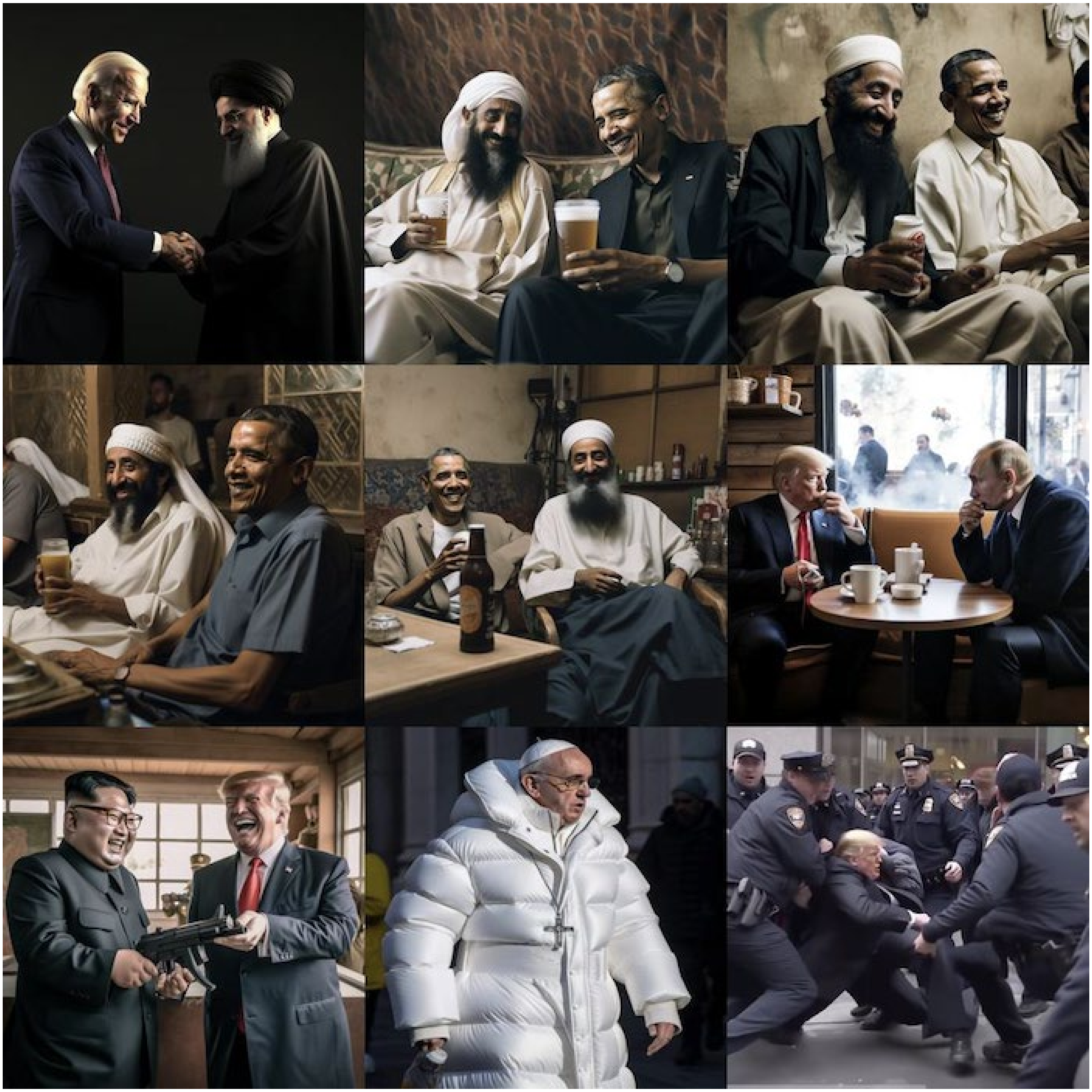

**Figure 1:** Synthetic images could be used for online election interference. Generated by Midjourney.

In our review, we identified several key case studies that illustrate the impact of GenAI on election interference (Chennupati, 2024; Kreps, *et al.*, 2022; Yang, *et al.*, 2019). For instance, during the 2016 U.S. presidential election, AI-generated botnets played a significant role in spreading misinformation and influencing public opinion (Bessi and Ferrara, 2016; Boichak, *et al.*, 2018; Howard, *et al.*, 2018; Shao, *et al.*, 2018). Similarly, deepfake videos have been used to create false narratives about political candidates, affecting voter perceptions and trust (Diakopoulos and Johnson, 2021; Dobber, *et al.*, 2021; Seymour, *et al.*, 2023; Vaccari and Chadwick, 2020). The deployment of GenAI in online election interference poses a unique challenge to democracies worldwide. Traditional methods of combating misinformation and digital manipulation are often insufficient to address the sophisticated techniques enabled by GenAI. This necessitates a comprehensive understanding of the various ways in which GenAI can be exploited for nefarious purposes and the development of new strategies to mitigate these risks (Gupta, *et al.*, 2023; Xu, *et al.*, 2023; Ziems, *et al.*, 2023).

This perspective article aims to shed light on the darker applications of GenAI in the context of online election interference. By examining the various nefarious uses of these technologies, we hope to highlight the urgent need for regulatory oversight, technological solutions, public awareness, and collaborative efforts to safeguard the integrity of democratic processes. Through this exploration, we seek to provide a foundation for policy-makers, technologists, and civil society to develop robust strategies to protect elections from AI-driven interference, ensuring that the benefits of GenAI are not overshadowed by its potential for harm (Chang, *et al.*, 2021 Rudin, 2019).

*The dual nature of generative AI.* GenAI and LLMs, celebrated for their ability to process natural language and generate multimodal content, offer numerous benefits, including enhancing creativity and solving complex problems (Brown, *et al.*, 2020; OpenAI, *et al.*, 2023; Shen, *et al.*, 2024; Yao, *et al.*, 2023). These technologies have enabled advancements in various fields, from natural language understanding and translation to creative industries such as art and music (Cao, *et al.*, 2023; Franceschelli and Musolesi, 2024). However, these same capabilities can be exploited for malicious purposes, such as creating deepfakes, synthetic identities, and targeted misinformation campaigns (Barrett, *et al.*, 2023; Kang, *et al.*, 2024; Mozes, *et al.*, 2023).

On the positive side, GenAI can assist in drafting documents, writing code, generating realistic images and videos, and even conducting complex data analysis (Bail, 2024; Li, *et al.*, 2024; Mandapuram, *et al.*, 2018). These applications have immense potential to boost productivity and innovation across different sectors (Kanbach, *et al.*, 2024; Ooi, *et al.*, 2025). For example, AI-generated content can support marketing efforts by creating personalized advertisements or generating engaging social media posts. In healthcare, GenAI can help in diagnosing diseases by analyzing medical images or generating detailed patient reports (Davenport and Kalakota, 2019; Sai, *et al.*, 2024).

Despite these benefits, the dark side of GenAI cannot be ignored (Kang, *et al.*, 2024; Wach, *et al.*, 2023). The same technology that generates realistic images and videos for legitimate purposes can also create deepfakes, which are manipulated media that convincingly alter appearances or actions of individuals. Deepfakes can be used to spread false information, damage reputations, and even blackmail individuals (Diakopoulos and Johnson, 2021; Seymour, *et al.*, 2023; Vaccari and Chadwick, 2020). The ease with which these deepfakes can be created poses a significant threat to the authenticity of information shared online. Similarly, LLMs can be used to generate convincing fake news articles, social media posts, and even entire personas (Augenstein, *et al.*, 2024; Chen and Shu, 2024; Choi and Ferrara, 2024a, 2024b; Quelle and Bovet, 2024). These synthetic identities can be employed to infiltrate social networks, manipulate public opinion, and disrupt democratic processes (Haider, *et al.*, 2023; Luceri, Boniardi, *et al.*, 2024). AI-generated text can mimic human writing so effectively that distinguishing between genuine and fake content becomes challenging (Augenstein, *et al.*, 2024; Jones and Bergen, 2024; Quelle and Bovet, 2024; Zellers, *et al.*, 2019). Moreover, GenAI's ability to tailor messages to specific demographics makes it a powerful tool for targeted misinformation campaigns. These campaigns can be designed to exploit societal divisions, amplify existing biases, and influence voter behavior (Boichak, *et al.*, 2018; Bovet and Makse, 2019; Ezzeddine, *et al.*, 2023; Goldstein, *et al.*, 2023; Hofmann, *et al.*, 2024; Sap, *et al.*, 2019; Ye, *et al.*, 2024). The targeted nature of these campaigns increases their effectiveness and the potential harm that they can cause.

The dual nature of GenAI highlights the need for a balanced approach in leveraging these technologies. While the benefits are substantial, the risks are equally significant (Barrett, *et al.*, 2023; Ferrara, 2024a; Weidinger, *et al.*, 2022, 2021). As we continue to develop and deploy GenAI, it is crucial to implement safeguards that prevent its misuse (Rudin, 2019). This includes developing robust detection mechanisms, establishing ethical guidelines, and promoting public awareness about the potential dangers of AI-generated content (European Commission. High-Level Expert Group on Artificial Intelligence, 2019; Jobin, *et al.*, 2019; Mittelstadt, *et al.*, 2016; Qiu, *et al.*, 2023; Shahriari and Shahriari, 2017; Sison, *et al.*, 2024; Smuha, 2019; Yang, *et al.*, 2019; Zhou, *et al.*, 2024). In conclusion, while GenAI holds great promise for enhancing various aspects of society, it also presents substantial risks. Understanding the dual nature of these technologies is essential for developing strategies that maximize their benefits while mitigating their potential harms (Mozes, *et al.*, 2023; Seymour, *et al.*, 2023).

*Survey approach*. To assess the growing body of scholarly work on the risks of generative AI (GenAI) in the context of online election interference, we conducted a systematic literature review. This review focused on identifying empirical studies, theoretical analyses, and case reports that examine the nefarious applications of GenAI technologies in political contexts. Our goal was to synthesize key trends, highlight influential contributions, and surface gaps in the existing research. We searched major academic databases — Google Scholar, Scopus, and IEEE Xplore — using targeted keywords such as “generative AI,” “deepfakes,” “LLMs,” “election interference,” “misinformation,” and “synthetic media.” The review spanned publications from 2018 through early 2025 and included peer-reviewed articles, preprints, government and NGO reports, and selected white papers. To ensure quality and relevance, we included only sources that offered clear empirical findings or policy-oriented analysis. We excluded works focused solely on AI development without direct relevance to election interference. This structured approach allowed us to capture a diverse and representative picture of GenAI’s evolving threat landscape.

## Nefarious applications in election interference

The integration of GenAI into the sphere of online election interference introduces numerous sophisticated threats to the integrity of democratic processes. These applications leverage the advanced capabilities of AI to manipulate information and influence voter behavior in unprecedented ways (Ferrara, *et al.*, 2020). Table 1 shows the primary nefarious applications of GenAI in election interference.

**Table 1: Summary of nefarious applications of GenAI in online election interference.**

| Nefarious application | Mechanism | Impact | Examples | Countermeasures | Future trends |
|---|---|---|---|---|---|
| **Deepfakes** | Erosion of public trust, reputation damage | AI-generated realistic videos/audio | Fake celebrity endorsements, political smear campaigns | AI-based detection tools, legal regulations | More sophisticated, harder-to-detect deepfakes |
| **AI-powered botnets** | Automated social media accounts | Distortion of public opinion, misinformation spread | Bot campaigns during elections, coordinated disinformation | Bot detection algorithms, platform policies | Increasingly human-like behavior, decentralized bot networks |
| **Targeted misinformation campaigns** | Tailored false content | Exploitation of societal divisions, voter manipulation | False news articles, doctored images | Fact-checking services, user education | AI-generated personalized misinformation |
| | | Infiltration of | Fake profiles | | More |
| **Synthetic identities** | AI-created personas | online communities, intelligence gathering | on social media, false expert identities | Identity verification protocols, AI detection tools | convincing and diverse synthetic identities |

***Deepfakes and synthetic media***

Among the most alarming applications of generative AI (GenAI) is the creation of deepfakes and other synthetic media that convincingly mimic real individuals and events. Using technologies like generative adversarial networks (GANs), AI systems can fabricate highly realistic videos, audio recordings, and images — often depicting public figures saying or doing things they never actually did (Seymour, *et al.*, 2023). These false representations are increasingly deployed to spread disinformation, damage reputations, and manipulate political narratives.

The implications for democratic processes are profound. A single deepfake video showing a political candidate engaging in unethical behavior or making inflammatory remarks could become viral and influence voter perception before it can be effectively debunked. Social media platforms, with their speed and scale of dissemination, further exacerbate this challenge (Pinto, *et al.*, 2025; Wach, *et al.*, 2023). A notable real-world example involved a fabricated video of Ukrainian President Volodymyr Zelenskyy urging troops to surrender — swiftly proven false but widely circulated before corrections could reach many viewers.

Beyond video, synthetic audio poses a similarly insidious threat. Voice cloning tools can replicate tone, accent, and speech patterns, creating fake soundbites that are nearly indistinguishable from authentic recordings. Likewise, AI-generated images of fictitious people or events have been used to provoke public outrage or fabricate visual "evidence" for political claims.

It is crucial to distinguish these perceptual threats—rooted in visual and auditory realism — from text-based threats produced by large language models (LLMs). While LLMs can generate persuasive and scalable misinformation, synthetic media leverages sensory cues that typically elicit higher trust, making their influence especially potent. Text-based disinformation tends to be more subtle and diffuse, whereas synthetic media often delivers sharper, more visceral impacts.

Recognizing these modality-specific risks is essential for developing effective countermeasures. Tailored strategies must account for the unique properties of visual, audio, and text-based GenAI outputs — guiding both the design of detection technologies and the formulation of policy responses.

***Botnets and social media manipulation***

AI-powered botnets represent another potent tool for election interference. Botnets are networks of automated accounts that can post, like, share, and comment on social media platforms, creating the illusion of widespread support or opposition for certain viewpoints (Yang and Menczer, 2023). These bots can be programmed to spread misinformation, amplify divisive content, and suppress legitimate discourse, thereby distorting the digital public sphere (Ferrara, *et al.*, 2016).

The sophistication of modern AI allows these bots to mimic human behavior convincingly, making it challenging to detect and eliminate them (Ferrara, 2023a). For example, during the 2016 U.S. presidential election, bots were used extensively to influence online discussions and spread false information, contributing to the overall misinformation ecosystem (Ferrara, 2018; Ferrara, *et al.*, 2016). Pozzana and Ferrara (2020) provided insights into the dynamics between bot and human behaviors, illustrating the increasing sophistication of bot activities; Luceri (2019a,b) showed how these behaviors evolve as a function of political ideology. The use of such technology in future elections poses a continuing threat to the integrity of democratic processes (Yang, *et al.*, 2022).

***Targeted misinformation campaigns***

GenAI's ability to synthesize realistic and persuasive text enables the creation of highly effective targeted misinformation campaigns. These campaigns are designed to exploit existing societal divisions and biases, tailoring messages to specific demographic groups to maximize their impact (Vosoughi, *et al.*, 2018). By leveraging data from social media platforms and other online sources, malicious actors can create personalized propaganda that resonates deeply with targeted individuals, influencing their opinions, emotions, and behaviors (Blas, *et al.*, 2024).

For example, during an election, a targeted misinformation campaign might disseminate false information about a candidate's policies or personal life, aiming to dissuade certain voter groups from supporting them. The personalized nature of these messages makes them particularly effective, as they can address the specific concerns and biases of the recipients (Allcott and Gentzkow, 2017). This form of manipulation undermines the democratic process by skewing the electorate's perception based on falsehoods.

***Synthetic identities and fake accounts***

GenAI can also be used to create convincing synthetic identities and fake accounts, which can be employed to infiltrate online communities, spread misinformation, and gather intelligence on political opponents (Ferrara, 2023a, 2024b). These synthetic identities can come with complete profiles, including photos, bios, and posting histories, making them difficult to distinguish from real users (Ezzeddine, *et al.*, 2023; Westerlund, 2019). The creation of fake accounts can facilitate a range of malicious activities. For instance, these accounts can be used to spread false narratives, engage in coordinated attacks against political figures, or even influence online polls (Cinus, *et al.*, 2024; Minici, *et al.*, 2025, 2024). The use of synthetic identities to manipulate online discourse presents a significant challenge for social media platforms and requires sophisticated detection and countermeasures (Haider, *et al.*, 2023).

In summary, the nefarious applications of GenAI in election interference are diverse and highly sophisticated. From creating deepfakes to orchestrating botnets, these technologies offer malicious actors powerful tools to disrupt democratic processes and influence voter behavior (Balasubramanian, *et al.*, 2023). Addressing these threats requires a multifaceted approach that includes technological solutions, regulatory measures, and public awareness initiatives (Ferrara, 2024b).

## Societal implications

Election interference poses several significant societal risks. These risks extend beyond the immediate impact on elections (see Table 2) and their outcomes and have profound implications for democratic institutions, social cohesion, and public trust (Badawy, Addawood, *et al.*, 2019; Broniatowski, *et al.*, 2018; Chang, *et al.*, 2021; Im, *et al.*, 2020; Ratkiewicz, *et al.*, 2011; Stella, *et al.*, 2018).

**Table 2: Examples of past election interference operations enabled by AI or bots.**

| Election | Country | Techniques used | Actors involved | Outcome | References |
|---|---|---|---|---|---|
| | | Bots, social | Pro-Brexit | Influenced public opinion, | Howard |
| 2016 | U.K. | media campaigns, fake news | groups, foreign actors | supported Brexit campaign | and Kollanyi, 2016 |
| 2016 | U.S. | Botnets, fake news, social media manipulation | Russian state-sponsored groups | Influenced public opinion, increased polarization | Ferrara, *et al.*, 2020; Vosoughi, *et al.*, 2018 |
| 2016 | Philippines | Troll accounts, fake news production | Domestic political actors, misinformation agents | Manipulated public opinion, supported political candidates | Haider, *et al.*, 2023; Ong and Cabañes, 2018 |
| 2017 | France | Deepfakes, hacked e-mail messages, social media campaigns | Russian state-sponsored groups, domestic actors | Attempted to discredit candidates, disrupted public discourse | Ferrara, 2017; Pierri, *et al.*, 2023 |
| 2017 | Spain | Bots, fake news,social media manipulation | Pro-separatist groups, foreign actors | Influenced public opinion, supported Catalan independence | Stella, *et al.*, 2018 |
| 2017 | Germany | Bots, trolls, misinformation | Russian state-sponsored groups | Influenced public opinion, attempted to disrupt political stability | Schwirtz and Borgia, 2019 |
| 2018 | U.S. | Social media bots, fake news, targeted ads | Foreign actors, domestic political groups | Influenced voter behavior, increased polarization | Deb, *et al.*, 2019 |
| 2014/16 & 2018/22 | Brazil | Social media bots, fake news, WhatsApp campaigns | Domestic political groups | Polarized electorate, influenced election results | Arnaudo, 2017; Pacheco, 2024 |
| 2018 | Italy | Social media bots, misinformation campaigns | Various political actors, misinformation agents | Influenced public opinion, affected election outcomes | Caldarelli, *et al.*, 2020. |
| 2019 | Indonesia | Spear phishing, cyber attacks, social media manipulation | Various political actors, cyber agents | Influenced voter behavior, disrupted electoral process | Bossetta, 2018 |
| | | Fake social | Various | Spread | |
| 2019 | India and Pakistan | media accounts, coordinated misinformation | political parties, independent actors | misinformation, manipulated voter sentiment | Banaji and Bhat, 2018 |
| 2020 | U.S. | Deepfakes, AI-generated fake news, targeted ads | Foreign actors, political groups | Increased misinformation, voter suppression efforts | Ferrara, 2020; Ferrara, *et al.*, 2020 |
| 2020 | Taiwan | Digital civic participation, misinformation campaigns | Various political actors, misinformation agents | Spread of misinformation, influence on voter perception and behavior | Chang, *et al.*, 2021 |

In Tables 3 and 4 we present a typology-based overview of GenAI threats — organized by modality, target, objective, and use case, respectively for democratic and authoritarian contexts.

**Table 3: Typology of GenAI applications in democratic contexts.**

| Modality | Target | Objective | Example use case |
|---|---|---|---|
| Text (LLMs) | Voters | Disinformation | Mass-produced false narratives on social media (Weidinger, *et al.*, 2021) |
| Text (LLMs) | Institutions | Disruption | AI-generated official-looking announcements or ballots (Chiacchiaro, 2025) |
| Audio | Voters | Discrediting | Spoofed robocalls with fabricated candidate statements (Schick, *et al.*, 2023) |
| | | | Fake videos of misconduct during election |
| Video (Deepfakes) | Candidates | Discrediting | campaigns (Marquart, 2023; Seymour, *et al.*, 2023) |
| Memes (Image+Text) | Voters | Polarization | Viral memes spreading divisive misinformation (Chen, *et al.*, 2025; Pinto, *et al.*, 2024) |
| Chatbots | Voters | Manipulation | Impersonating campaign or electoral info services (Chiacchiaro, 2025; Li, *et al.*, 2024) |

**Table 4: Typology of GenAI applications in authoritarian contexts.**

| Modality | Target | Objective | Example use case |
|---|---|---|---|
| Image (GANs) | Voters | Amplification | Synthetic photos used in pro-regime propaganda (Farrell, *et al.*, 2022; Gilding, 2023) |
| Video (Deepfakes) | Citizens | Disruption | Fabricated dissent to justify crackdowns (Hameleers and Marquart, 2023; Le, *et al.*, 2022) |
| Text | | | Ghostwritten op-eds simulating grassroots |
| (LLMs) | Institutions | Disinformation | support (McGuffie and Newhouse, 2020) |
| Interactive media | Voters | Indoctrination | Game-like simulations promoting regime narratives (Kidd and Birhane, 2023) |

***Erosion of public trust***

The widespread use of deepfakes and misinformation can severely erode public trust in media, institutions, and the democratic process itself. When voters are exposed to manipulated content that appears authentic, their ability to discern truth from falsehood is compromised. This erosion of trust can lead to cynicism and apathy among the electorate, as people become increasingly skeptical of all information sources (Seymour, *et al.*, 2023). The impact of this distrust is far-reaching. It undermines the credibility of legitimate news organizations and government institutions, making it easier for malicious actors to spread further misinformation (Baribi-Bartov, *et al.*, 2024). Once trust is broken, restoring it is complex, requiring significant efforts from involved stakeholders like media organizations, governments, and broader society (Ferrara, 2024b).

***Polarization and division***

Targeted misinformation campaigns exacerbate existing societal divisions by exploiting and amplifying biases and prejudices. By delivering tailored messages that resonate with specific demographic groups, these campaigns can deepen political and social polarization. For instance, misinformation targeting different ethnic or social groups can inflame tensions and lead to increased hostility between communities (Vosoughi, *et al.*, 2018). Such polarization weakens the social fabric, making it more challenging to achieve consensus on critical issues. This division can manifest in various forms, from increased partisanship in political discourse to violent confrontations between opposing groups. The long-term consequences of such division are detrimental to social stability and cohesion (Allcott and Gentzkow, 2017).

***Undermining democracy***

The ability to manipulate public opinion on a large scale poses a direct threat to the foundations of democracy. Elections are meant to reflect the informed choices of the electorate. However, when these choices are influenced by false information, the legitimacy of the electoral process is compromised (Luceri, Ye, *et al.*, 2024). This manipulation undermines the principle of free and fair elections, a cornerstone of democratic governance (Vosoughi, *et al.*, 2018). Badawy, Lerman, *et al.* (2019) examined the demographics of individuals most susceptible to online political manipulation, highlighting the vulnerability of certain groups.

Furthermore, the perception that elections can be easily manipulated through AI-driven misinformation campaigns can lead to decreased voter turnout. If people believe their votes do not matter or that the process is rigged, they are less likely to participate in the democratic process. This disengagement weakens democratic institutions and can lead to governance that is less representative of the people’s will (Ferrara, 2023a).

### *Exacerbation of inequality*

The deployment of GenAI technologies in ways that disproportionately affect certain groups can exacerbate existing social inequalities (Ferrara, 2023b). For example, misinformation campaigns might target marginalized communities with content designed to suppress their voter turnout or sow confusion about voting processes (Calo, 2017). Such tactics exploit the vulnerabilities of these communities, further entrenching their disenfranchisement. Moreover, access to sophisticated AI technologies is often concentrated among well-resourced entities, such as state actors or large corporations, giving them disproportionate power to influence public opinion and elections (Pinto, *et al.*, 2024). This concentration of power can lead to a democratic deficit, where the interests of a few overshadow the voices of the many (Goldstein, *et al.*, 2023).

### *Psychological impact on society*

The constant exposure to AI-generated misinformation and deepfakes can have a psychological toll on the populace. The bombardment of conflicting information can lead to information overload, where individuals struggle to process and make sense of the vast amounts of data they encounter daily: This situation can cause increased stress, anxiety, and a sense of helplessness among individuals (U.S. Department of Homeland Security, 2014; Fu, *et al.*, 2020; Matthes, *et al.*, 2020). Additionally, the fear of being deceived by AI-generated content can lead to a general mistrust of digital information. This mistrust can hinder the adoption of beneficial technologies and stifle innovation, as people become wary of engaging with digital platforms (Ferrara, 2024b).

### *Prevalence, moral panic, and regulatory targets*

While concern over GenAI threats has grown significantly, the actual number of confirmed cases of election interference remains limited. This discrepancy raises the question of whether current discourse may be partially shaped by moral panic—the tendency to overestimate the immediate risks of novel technologies.

That said, GenAI is uniquely positioned to amplify threats at scale and speed, which justifies careful preemptive policy discussions. Vulnerable populations — such as older adults, individuals with low media literacy, or politically disengaged users — are more susceptible to synthetic content (Mandapuram, *et al.*, 2018; Ye, Luceri, *et al.*, 2024). Their susceptibility makes them not only targets but potential amplifiers of GenAI misinformation.

From a policy perspective, regulatory frameworks should consider not only content creators and platforms but also the pathways of dissemination and the psychological traits that influence gullibility. Educational initiatives focused on digital literacy, platform accountability, and cross-sector collaboration will be vital in building long-term societal resilience.

### *Applications in authoritarian contexts*

While the majority of documented GenAI election interference campaigns are observed in democratic or quasi-democratic contexts, the potential for exploitation in authoritarian regimes is substantial. However, empirical studies focused on authoritarian settings remain sparse—likely due to challenges in access, censorship, and the lack of media transparency.

Authoritarian governments may use GenAI not only to disrupt external democratic processes but also to maintain internal control. Tactics may include generating synthetic dissent to justify crackdowns, deploying AI-generated propaganda to simulate mass support, or discrediting foreign critics through deepfakes and forged documents.

The opacity of such regimes makes it difficult to quantify the scope and prevalence of GenAI use. This gap in the literature is a critical limitation, and future research must find creative and safe methodologies to illuminate the uses and abuses of AI technologies in these environments.

## Mitigation strategies

To combat the multifaceted threats posed by GenAI in online election interference, a comprehensive and multi-pronged approach is essential. This section outlines various strategies, including regulatory measures, technological solutions, public awareness campaigns, and collaborative efforts, aimed at mitigating the risks associated with the misuse of GenAI.

### *Regulation and oversight*

Effective regulation and oversight are crucial in governing the use of GenAI, particularly in sensitive areas like elections. Governments and international bodies need to establish clear guidelines and standards for the ethical use of AI technologies. These regulations should include provisions for transparency, accountability, and the prevention of misuse (Floridi, 2019).

One approach is the implementation of AI ethics guidelines, such as the European Union's "Ethics guidelines for trustworthy AI", which emphasize principles like human agency, technical robustness, privacy, and accountability (European Commission. High-Level Expert Group on Artificial Intelligence, 2019). Additionally, regulatory frameworks should mandate the disclosure of AI-generated content to ensure that the public can identify and scrutinize such content appropriately (Solaiman, *et al.*, 2019).

### *Technological solutions*

Advances in technology can provide robust solutions to detect and mitigate the impact of AI-generated misinformation. For instance, digital watermarking and forensic techniques can be employed to identify deepfakes and other synthetic media (Perov, *et al.*, 2020). These technologies can help verify the authenticity of digital content, making it easier to distinguish genuine from manipulated media.

Moreover, AI-driven detection systems can be developed to identify and neutralize botnets and fake accounts on social media platforms. These systems can use machine learning algorithms to detect patterns indicative of automated behavior, enabling platform operators to take swift action against malicious bots (Yang, *et al.*, 2019). Implementing such technologies requires collaboration between tech companies, researchers, and policy-makers to ensure their effectiveness and scalability.

In response to the rising sophistication of GenAI, the research community has made progress in developing detection tools. For example, Salvi, *et al.* (2023) introduced the TIMIT-TTS dataset, a benchmark for evaluating text-to-speech detection methods, which helps advance audio-based synthetic media detection. Similarly, Hosler, *et al.* (2021) proposed a semantic approach to deepfake detection that leverages emotional inconsistencies, offering a promising direction for identifying AI-generated visual content. These works underscore the importance of aligning detection technologies with the specific modality of GenAI content. As generative systems continue to evolve, multimodal detection techniques and adaptive classifiers will be crucial components of an effective defense strategy.

### *Public awareness and education*

Educating the public about the potential for AI-driven misinformation and how to critically evaluate information sources is crucial in combating the spread of false information. Public awareness campaigns can help individuals develop the skills needed to identify deepfakes, recognize signs of bot activity, and verify the credibility of news sources (Rudin, 2019).

Educational initiatives should be integrated into school curricula and adult education programs to promote media literacy. These programs can teach critical thinking skills and provide tools for assessing the reliability of information online (Ferrara, 2023a). Additionally, public service announcements and partnerships with media organizations can amplify these efforts, reaching a broader audience.

### *Collaborative efforts*

Addressing the challenges posed by GenAI requires collaborative efforts among policy-makers, technologists, and civil society. Multi-stakeholder initiatives can foster the development of best practices and shared standards for the ethical use of AI. For example, the Partnership on AI brings together diverse organizations to advance public understanding of AI and promote the responsible use of AI technologies (Heer, 2018).

Moreover, international cooperation is essential to address the global nature of online election interference. Governments and international organizations should work together to share information, coordinate responses, and develop joint strategies to counteract AI-driven threats. This collaboration can help build a unified front against the misuse of GenAI, ensuring that efforts are aligned and resources are effectively utilized (Taddeo and Floridi, 2018).

### *Ethical AI development*

Promoting ethical AI development involves ensuring that AI systems are designed with fairness, accountability, and transparency in mind. Developers and researchers should adhere to ethical guidelines and consider the societal impact of their work. This includes conducting thorough risk assessments, implementing bias mitigation strategies, and fostering inclusive design practices (Weidinger, *et al.*, 2022).

Ethical AI development also involves creating AI systems that are interpretable and explainable. Interpretable models allow users to understand how decisions are made, increasing trust in AI systems and enabling more effective oversight (Rudin, 2019). By prioritizing ethical considerations throughout the AI development lifecycle, developers can help prevent the misuse of AI technologies in election interference and other contexts (Bellamy, *et al.*, 2019).

### *Legal and policy frameworks*

Developing robust legal and policy frameworks is essential to address the challenges posed by GenAI. These frameworks should encompass data protection laws, cybersecurity measures, and regulations specific to AI technologies. For instance, the General Data Protection Regulation (GDPR) in the European Union provides a strong foundation for data privacy and security, which is crucial in preventing unauthorized access and manipulation of data used in AI systems (Goodman and Flaxman, 2017).

Policies should also address the accountability of AI developers and users, ensuring that there are consequences for the misuse of AI technologies. This can include penalties for creating and disseminating deepfakes, as well as requirements for transparency in AI-generated content. Legal frameworks need to be adaptive and responsive to the rapidly evolving AI landscape, ensuring they remain effective in mitigating emerging threats (Mittelstadt, *et al.*, 2016).

In conclusion, addressing the societal implications of GenAI in election interference requires a comprehensive approach that integrates regulatory measures, technological solutions, public education, collaborative efforts, ethical development practices, and robust legal frameworks. By implementing these strategies, we can mitigate the risks associated with GenAI and protect the integrity of democratic processes (Ferrara, 2024a).

## Conclusions

This study highlights the significant risks posed by generative artificial intelligence (GenAI) in the context of online election interference. By employing a dual methodology of literature review and analysis, we have provided a comprehensive overview of the nefarious applications of GenAI. Our findings underscore the urgent need for robust mitigation strategies, including regulatory measures, technological solutions, public awareness campaigns, and international cooperation.

Our research contributes to the existing body of knowledge by elucidating the specific threats posed by GenAI and proposing actionable recommendations to safeguard democratic processes. Future research should focus on developing advanced detection mechanisms and exploring the ethical implications of GenAI to ensure its responsible use.

## About the author

Emilio Ferrara is a professor of computer science and communication in the Thomas Lord Department of Computer Science at the University of Southern California.
E-mail: emiliofe [at] usc [dot] edu

## About the team

The 2024 Election Integrity Initiative is led by Emilio Ferrara and Luca Luceri and carried out by a collective of USC students and volunteers whose contributions are instrumental to enable these studies. The author is grateful to the following HUMANS Lab's members for their tireless efforts on this project: Ashwin Balasubramanian, Leonardo Blas, Charles 'Duke' Bickham, Keith Burghardt, Sneha Chawan, Vishal Reddy Chintham, Eun Cheol Choi, Srilatha Dama, Priyanka Dey, Isabel Epistelomogi, Saborni Kundu, Grace Li, Richard Peng, Gabriela Pinto, Jinhu Qi, Ameen Qureshi, Namratha Sairam, Tanishq Salkar, Srivarshan Selvaraj, Kashish Atit Shah, Gokulraj Varatharajan, Reuben Varghese, Siyi Zhou, and Vito Zou.

## Further readings

(F) *New York Times* — "Do these A.I.-created fake people look real to you?" (21 November 2020), at https://www.nytimes.com/interactive/2020/11/21/science/artificial-intelligence-fake-people-faces.html, accessed 29 April 2025.

(G) *Marketplace* — D. Brancaccio, A. Schroeder, and E. Soderstrom, 2023. "AI amplifies scam calls and other deceptions" (14 July), at https://www.marketplace.org/story/2023/07/14/ai-amplifies-scam-calls-and-other-deceptions, accessed 29 April 2025.

(H) *CBS News* — C. Evans and A. Novak, 2023. "Scammers use AI to mimic voices of loved ones in distress" (19 July), at https://www.cbsnews.com/news/scammers-ai-mimic-voices-loved-ones-in-distress/, accessed 29 April 2025.

(I) *New York Times* — T. Hsu and S.A. Thompson, 2023. "Disinformation researchers raise alarms about A.I. chatbots" (8 February), at https://www.nytimes.com/2023/02/08/technology/ai-chatbots-disinformation.html, accessed 29 April 2025.

(J) *MIT Technology Review* — K. Hao, 2019. "The biggest threat of deepfakes isn't the deepfakes themselves" (10 October), at https://www.technologyreview.com/2019/10/10/132667/the-biggest-threat-of-deepfakes-isnt-the-deepfakes-themselves/, accessed 29 April 2025.

(K) *Reddit* — "r/midjourney," at https://www.reddit.com/r/midjourney/, accessed 29 April 2025.

(L) *NPR* — B. Allyn, 2022. "Deepfake video of Zelenskyy could be 'tip of the iceberg' in info war, experts warn" (16 March), at https://www.npr.org/2022/03/16/1087062648/deepfake-video-zelenskyy-experts-war-manipulation-ukraine-russia, accessed 29 April 2025.

---

---